\documentclass[dvips]{acta}
\usepackage{supertabular,lscape,epsfig}
\usepackage{amssymb}
\usepackage{amsmath}
\usepackage[T1]{fontenc}

\SetPages{0}{0}

\SetVol{76}{2026}

\usepackage{lmodern}

\begin{document}

\begin{Titlepage}

\Title{Boundary conditions for radial pulsations in AGB envelopes}

\Author{Zalewski, J.,}
{Independent researcher   \\
e-mail: jan.zalewski.a2@gmail.com}

\Received{May 4, 2026}
\end{Titlepage}

\Abstract{We present a formulation of outer and inner boundary conditions for radial non-adiabatic pulsations in AGB/post-AGB envelopes in terms of local structure of pulsation equations. This approach provides a framework to construct classes of boundary condition selectors.

We show that the pulsation spectrum is primarily determined by the choice of the outer boundary selector. Different choices of the two-dimensional subspace at the outer boundary lead to different types of pulsation spectra ranging from p-mode like spectra to ones including strongly excited and damped strange modes.

Our results show that the outer boundary conditions control the degree of coupling between acoustic and entropy components in the solution. Selectors dominated by slow (non-acoustic) branches lead to strongly non-adiabatic behavior and emergence of strongly excited/damped strange modes, while selectors including acoustic branches result in weaker coupling and more classical looking spectra.

At the inner boundary we found that it is sufficient to enforce that the solution resides in a two-dimensional subspace allowed by the local pulsation equations. This ensures regular behavior and compatibility with WKB asymptotics in deep parts of the envelope. Once this condition is met the detailed form of the inner boundary selector has only minor influence on the resulting pulsation spectrum.

We also compare the obtained spectra with those obtained using standard boundary conditions to show that the variations in excitation rates of strange modes reported previously may be interpreted in terms of the subspaces selected by the outer boundary.}

{stars: AGB and post-AGB, stars: oscillations, stars: interiors}

\section{Introduction}
Boundary conditions for pulsation are of particular importance in case of supergiant stars with extended envelopes as the transition from the envelope sustaining pulsations to the outside medium is gradual and does not offer a well defined border layers. The classical view of pulsation is a cavity surrounded on both inner and outer ends with either regularity conditions or allowing the use of algebraic relations between variables at the border, such as vanishing Lagrangian pressure perturbation at the outer boundary or relations resulting from linearized black body relation (Cox 1980, Christensen-Dalsgaard 2008). While such approach is justified for many types of stars it is not sufficient for supergiant stars and highly non-adiabatic pulsations.

Early analyses of giant and supergiant stars, and in particular the discovery by Wood (1976) of strange modes in helium stars have indicated that the pulsational properties in these stars depend on the adopted boundary conditions. Several approaches were developed over time to deal with the problem that pulsation region does not have fixed or well defined boundaries. These include asymptotic formulations enabling the inner boundary layer not to act as a rigid wall reflecting incoming waves (Dziembowski 1977, Osaki 1977). Later on multiple alternative formulations of outer boundary conditions were tried like fitting the envelope to atmosphere, using isothermal approximation or boundary conditions based on dispersion relations (see e.g. Christensen-Dalsgaard 2008, Townsend 1997, Saio 2010, also Zalewski 1991, 1992).

In spite of these efforts no single formulation of how to deal with boundaries in extended envelopes of supergiant stars has been adopted. This was attributed by Glatzel and Gautschy (1992) to the fact that the boundaries for model envelope do not correspond to physical boundaries of the star, which makes the boundary conditions non-unique. Saio et al. (1998) also discussed the influence of boundary conditions on strange modes in supergiant stars.  Aikawa (1991, 1993) has noted that selection of thermal boundary conditions affected the computed growth rates. In contrast the frequencies of pulsation and stability of normal p-modes in supergiants are much less affected by the choice of boundary conditions.

In what follows we are analyzing linear, radial, non-adiabatic pulsations in the envelopes of AGB/\allowbreak post-AGB stars for which normal and strange modes are found in pulsation studies. Our aim is a formulation of boundary conditions for both the inner as well as outer boundaries of the pulsation region and analysis of the effects that boundary condition selection has on pulsation. We also tie the formulation of boundary conditions to the method of iteration of eigenfrequencies in order to obtain a consistent and unified framework. 

A quantitative measure is introduced to estimate the effects of the selected boundary conditions on pulsations and to assess the obtained eigenmodes. Our formulation does not preclude the use of other forms of boundary conditions and in fact may be used to better understand the effects of such conditions on pulsation.

\section{Envelope Model and Pulsation Equations}

\subsection{Envelope models}
In what follows we study pulsational properties of post-AGB stars and therefore consider envelope models with \(M/M_{\odot}\leq 1\) and \(\log(L/L_{\odot})\sim 4\) (Paczy\'nski, 1970), and effective temperatures \(\log(T_{eff})=3.7 - 4.2\) covering the range where strange modes are found in these stars. In this paper we examine a model with \(M=0.69M_{\odot}\), \(L=10^{4}L_{\odot}\) and \(\log(T_{eff})=3.8 - 3.9\) which has been shown to exhibit strange modes for standard boundary conditions (see Zalewski, 1992) and thus provides as suitable model for the present study.

The surface radius \(R_{surf}\) and temperature \(T_{surf}\) are iterated so that black-body relation is satisfied at \(\tau=2/3\). In this way the meaning of effective temperature is made more consistent for the studied models of tenuous extended envelopes of AGB stars. 

The surface density of \(\log(\rho_{surf})=-12\) is typically assumed. The envelope models are integrated from the surface to a predefined maximum temperature \(T_{max} = 10^7K\). 

The opacities were taken from the OPAL tables of Iglesias and Rogers (1996), assuming the Grevesse and Noels (1993) heavy element mixture with hydrogen and metal abundances \(X=0.7\), \(Z=0.02\). At low temperatures, opacities were supplemented by the tables of Ferguson et al. (2005). A bi-cubic spline interpolation of opacity and its derivatives was used.

Convection in the envelope was treated using Paczy\'nski's (1969) formulation, which is suitable for the present case of linear pulsations in warmer supergiant envelopes. Dynamic effects of convection on pulsation were neglected (see Aikawa, 2010).

The envelope was integrated with a constant step in \(x=\ln(r/R_{\odot})\), typically of \(5\times 10^{-4}\). A Runge--Kutta (RK4) integration method was used.

\subsection{Pulsation models}
For linear non-adiabatic pulsations we use the same notation and dependent variables as in Dziembowski (1977). 

We use Cowling approximation, as the envelope mass in AGB models is negligible. We assume radial pulsation (\(l=0\)). 

The pulsation equations can thus be written as:
\begin{equation}
y'=M y
\label{eq:basicEq}
\end{equation}
with \(y=(d,p,s,f)\) being the dependent variables, and \(x\) being the independent variable. The coefficients of the matrix \(M\) are (see Dziembowski 1977):
\begin{equation}
M = 
\begin{bmatrix}
m_{11} & m_{12} & m_{13} & 0 \\
m_{21} & m_{22} & 0 & 0 \\
m_{31} & m_{32} & m_{33} & m_{34} \\
0 & 0 & m_{43} & 0
\end{bmatrix}
\label{eq:matrixM}
\end{equation}

In our case the \(m_{34}=1\), as luminosity is constant in our envelope models. We use non-dimensional \(\sigma\) related to pulsational frequency \(\omega\) as \(\omega=\sqrt{4 \pi G \langle\rho\rangle}\sigma\), with \(\exp({\omega t})\) time dependence.

Since the outer and inner boundaries of the envelope do not represent physical boundaries, the form of boundary conditions for pulsations has to be assumed. This will be discussed in Section 3.

\section{Formulation of Boundary Conditions as a Subspace Problem}

\subsection{Local eigenstructure of the pulsation matrix}

Since both the outer and inner boundaries in our models are placed in regions of the envelope far away from ionization zones and opacity bumps the matrix \(M\) exhibits a separation into a pair of large-magnitude eigenvalues with opposite signs and two smaller magnitude ones. This structure allows the solution to be locally decomposed into components associated with these distinct branches, which provides a basis for formulating boundary conditions. 

Near outer boundary of the envelope the branches of \(M\) are typically interpreted as corresponding to incoming and outgoing acoustic waves (see e.g. Mihalas and Mihalas, 1984), and to interpret the smaller magnitude eigenvalues as corresponding to non-acoustic branches. Likewise deep in the envelope, near the inner boundary eigenvalues of \(M\) also form a pair of large value \(\pm k\) and two of smaller magnitude. In this case the pair of large-magnitude eigenvalues \(\pm k\)  is associated with the thermal part of the solution, whose behavior, in the deep layers is dominated by the entropy perturbation (see Dziembowski, 1977).

Thus in both cases of boundaries distinct branches of \(M\) appear. And while typically the search for boundary conditions based on dispersion relations seeks to identify a specific kind of branch (like acoustic or thermal) we in what follows will only assume that it is possible to pick two of the four eigenvalues of \(M\) at each of the boundaries to obtain four conditions to satisfy the requirements of the fourth-order system of linear radial pulsation equations.

A method to achieve such selection at both boundaries is discussed in Section 4.  In the following subsections of Section 3 we will explore the implications of adopting the dispersion relation based approach to the formulation of boundary conditions.

The fact that at both boundaries the dominant eigenvalues occur in a pair differing by sign implies that there are two opposing branches of possible solutions (for the dominant component) - viz. one propagating away from the pulsation cavity and the other propagating towards the pulsation region. Customarily in case of the outer boundary conditions, where the dominant eigenvalues correspond to acoustic like perturbations, the eigenvalue corresponding to the propagating outwards perturbation or the one decaying outwards is adopted, while the complementary eigenmode is rejected on physical grounds. This implies that when considering which of the four eigenvalues and their corresponding eigenvectors to pick for boundary condition formulation one may postulate the existence of a two-dimensional subspace of \(\mathbb{C}^4\) containing eigenvectors of \(M\) that represent the flux leaving the envelope and a complementary 2D subspace containing those eigenvectors of \(M\) that correspond to the incoming flux from outside. 

Thus the search for boundary conditions may be viewed as a selection of 2D subspaces in which the solution of Eq.~(\ref{eq:basicEq}) should be contained near each boundary. We will refer to such a subspace as the 'wanted' one, with the complementary one being the unwanted subspace, which the solution \(y\) should avoid.

The current approach based on the local analysis of branches of pulsation matrix \(M\) to determine boundary conditions may be viewed as an alternative to the classically adopted algebraic boundary conditions relations (like \(p'=0\) or the linearized blackbody relation for outer boundary condition). We will examine this approach and its consequences for the formulation of boundary value problem for linear radial pulsations in the envelope in following subsections.

\subsection{Outer boundary conditions}

By selecting two of the four eigenvalues and their associated eigenvectors at the outer boundary it is possible to define the two linearly independent vectors:\(y_{s,i},\ i=1,2\) as:
\begin{equation}
	y_{s,i} = v_{s,i}
\label{eq:obcCond}
\end{equation}
with \(v_{s,i}\) being the eigenvectors of \( M_s v_{s,i}=k_{s,i} v_{s,i}\) corresponding to the selected two eigenvalues \(k_{s,i},\ i=1,2\) at the surface. 

These two vectors provide initial values for two independent solutions of Eq.~(\ref{eq:basicEq}).

Thus in our formulation, based on dispersion relations, the boundary values are constructed from the eigenvectors of \(M_{s}=M|_{surface}\), which define a two-dimensional subspace of admissible initial conditions. This form of boundary conditions specifies the full state vector at the boundary, unlike the situation with algebraic boundary conditions which impose relations between pulsational variables without necessarily defining all components of the solution.

\subsection{Integration of equations}

Several methods of integration of linear non-adiabatic pulsation equations, particularly in the presence of rapidly growing and decaying components, have been developed and they will not be discussed in this paper (see, e.g. Dziembowski 1977, Glatzel and Gautschy 1992, Townsend and Teitler 2013).

We are primarily using direct integration of Eq.~(\ref{eq:basicEq}) with a fourth order Runge--Kutta integration method, with the coefficients of matrix \(M\) linearly interpolated at the sub-grid points. 

While alternative approaches based on intermediate fitting points within the envelope exist, we integrate the equations to the inner boundary. The main requirement of the adopted integration method is to preserve the initial subspace, determined at the outer boundary, as it is propagated inward.

\subsection{Location of the inner boundary}

The envelope in our models extends from outer regions down to layers with temperature of \(10^7 K\). This covers the range used when computing linear non-adiabatic radial pulsations. In fact the typically adopted position of the inner boundary is related to temperature and ranges from \(10^6\) K to \(10^7\) K. The choice of the maximum temperature is affected by the stability of the used integration method as conventionally the computations were switched to WKB based approximations to avoid buildup of parasite components affecting the solution.

In our code we use a criterion based on the comparison of the dynamical time scale to the thermal time scale and place the inner boundary at a point where \(\tau_{th}/\tau_{dyn} \sim |\omega|\tau_{th} \gtrsim 1\). That is where the thermal time scale becomes comparable or exceeds the dynamical time scale. For fundamental radial mode this usually occurs around \(\log(T)\sim 6.8\), while for higher frequency modes the inner boundary layer would be located at somewhat lower temperatures, depending on the pulsation frequency. These are regions of the inner envelope where radial pulsations amplitude becomes exponentially small and both the kinetic energy \(EK\) of the mode and its dissipation integral \(ED\) are already stabilized well before the inner boundary is reached, thus all the mode energy and driving occurs in the envelope far from the boundary.

The position of the inner boundary in the envelope depends on the \(\omega\) and may vary during the iteration of pulsation frequency.  

\subsection{Inner boundary conditions}

Similarly to the case of selection of two eigenvalues at the outer boundary for the inner boundary, two eigenvalues (and their eigenvectors) are selected (see Section 4 for discussion of the selection process). These two eigenvalues are denoted as \(k_{b,i},\ i=3,4\) (the indexing scheme at inner boundary will become obvious below). Once the two eigenvalues are selected their corresponding eigenvectors \(v_{b,i},\ i=3,4\) span the 2D subspace which is expected to contain the solution of Eq.~(\ref{eq:basicEq}). The two remaining eigenvectors span the unwanted subspace.

As in the case of outer boundary, the subspace provides complete information about the expected state vectors at the inner boundary.

While the surface vectors of the solution of Eq.~(\ref{eq:basicEq}) automatically span the wanted subspace the integrated vectors \(y_{b,i},\ i=1,2\) need not span the wanted subspace at the inner boundary. In fact, unless the correct eigenfrequency of pulsation is used, the solutions of Eq.~(\ref{eq:basicEq}) at the inner boundary will have components in the unwanted subspace, and it is the process of finding proper pulsation frequency that eliminates the components of \(y_{b,i}\) from that subspace.

In case of non-normal matrix \(M\) it is possible to quantify the amount of unwanted components in the solution by using left eigenvectors of \(M\): \(w^{\dagger}_i M =k_i w^{\dagger}_i\). When normalized left eigenvectors \(w_{i},\ i=1,\ldots, 4\) satisfy \(w^{\dagger}_{i} v_{j} = \delta_{i,j}\). They can be used to determine the component of the solution vector along the corresponding eigenvector \(v_i\). If \(w_{b,i},\ i=1,2\) are the two left eigenvectors that correspond to the unwanted eigenvalues \(k_{b,i},\ i=1,2\) then
\begin{equation}
 b_{i,j} = w^{\dagger}_{b,i}\ y_{b,j},\ i,j=1,2 
\label{eq:bijDef}
\end{equation} 
measure the projection of the \(y_{b,j}\) vectors onto the unwanted subspace spanned by \(v_{b,i}\) eigenvectors.

Eq.~(\ref{eq:bijDef}) may be seen as the formulation of the usual pulsational matrix 
\begin{equation}
 B=(b_{i,j})_{i,j=1,2}
 \label{eq:matrixB}
\end{equation} expressed in a form suitable for non-normal systems.

It may be concluded that the formulation of both the outer and inner boundary conditions in terms of the expected 2D subspaces spanned by the solution at both ends provides complete state vectors and enables a proper interpretation of the matrix \(B\). The current formulation provides a framework for controlling the behavior of the solution near the boundaries and may improve the accuracy of the obtained eigenfrequencies and eigenmodes.
 
\subsection{Eigenfrequency condition}

Since it is not the individual components \(y_{i},\ i=1,2\) of the solution of Eq.~(\ref{eq:basicEq}) that are required to vanish in the unwanted subspace  at the inner boundary, but rather the solution \(y = c_1 y_1+c_2 y_2\) formed as their linear combination, the condition may be expressed as \(w^{\dagger}_i y = 0, \ i=1,2\). Therefore the condition that determines the eigenfrequency is not for the individual coefficients \(b_{i,j}\) to vanish but rather that the determinant of \(B\) in Eq.~(\ref{eq:matrixB}) be zero:
\begin{equation}
	\det(B) = 0
\label{eq:detB}
\end{equation}

This means that the determinant vanishes when there exists a non-trivial linear combination that satisfies the boundary conditions. 

In the present case the determinant plays the same role as in the traditional formulation and is used to iterate the frequency of pulsation using Newton--Raphson method. With the present formulation the convergence rate is fast, typically several iterations are required. Thus, the adopted approach does not substantially differ as far as the rate of convergence or the domain of convergence are concerned from conventional cases.

\subsection{Reconstruction of the eigenmode}

In order to reconstruct the solution of Eq.~(\ref{eq:basicEq}) from its components \(y_i,\ i=1,2\), once the frequency of pulsation is found, matrix \(B\) is used because the coefficients of linear combination of \(y_i\) satisfy:
\begin{equation}
	B c = 0
\label{eq:condC}
\end{equation}
which follows from \(w^{\dagger}_i\ y = c_1\ w^{\dagger}_i\ y_1+c_2\ w^{\dagger}_i\ y_2 = 0, \ i=1,2\). One of the \(c_i\) is set to 1 and the other is determined from Eq.~(\ref{eq:condC}). 

The mode thus reconstructed has the property that at each boundary it lies completely in the wanted subspaces, thus the state vector does not have any components within the unwanted subspaces at the boundaries. 

\subsection{Quantitative diagnostics}

From the derivation presented in the preceding subsections it follows that our approach may be used to quantify how well the obtained solution satisfies the assumed boundary conditions by measuring the projection of the solution vector onto the unwanted subspaces at each boundary. 

A suitable measure \(\eta\) is obtained in the following manner. The projection of the solution at the boundary \(y\) onto the four eigenvectors of \(M\) is obtained from:
\begin{equation}
	\gamma_{i} = \frac{w^{\dagger}_i\ y}{w^{\dagger}_i v_i}
\label{eq:defGamma}
\end{equation}
where \(i=1,\ldots,4\). By forming ratios of the \(\gamma_i\), representing projections onto the two unwanted eigenvectors with the corresponding \(\gamma_j\) for the wanted eigenvectors two values of \(\eta_i,\ i=1,2\) at each boundary are obtained:
\begin{equation}
	\begin{aligned}
\eta_1 &= \gamma_1/\gamma_3 \\
\eta_2 &= \gamma_2/\gamma_4
	\end{aligned}
\label{eq:defEtas}	
\end{equation}
where \(\gamma\) indices \(i=1,2\) correspond to unwanted components and \(i=3,4\) to the wanted subspace.

These \(\eta\)'s represent the contamination of the solution at the boundary originating from the component of the solution vector lying in the unwanted subspace.

If any of \(|\eta_i|,\ i=1,2\) becomes large it indicates that the obtained eigenmode is contaminated by the contribution from the unwanted subspace. It may be a signal of a numeric problem with the solution and is one of the indicators we use to verify the obtained pulsation modes.

If the eigenmode is obtained using other formulation of boundary conditions the approach presented here could still be used to obtain useful information on which branches have the largest contribution to the adopted boundary conditions.

\section{Boundary-condition selection and its effects}

In this section we classify the types of boundary conditions resulting from the dispersion-relation-based formulation and examine their effects on radial pulsations in AGB envelopes. Since neither the physical conditions near boundaries nor the dispersion-relation-based formulation uniquely determine their selection, it is therefore useful to examine systematically how different admissible choices affect the resulting pulsation spectrum and eigenfunctions. This will be done by examining the properties of the resulting pulsation, such as eigenfrequency spectra and the behavior of pulsation eigenmodes.

\subsection{Classification of boundary-condition selectors}

As discussed in Section 3, near both the outer and inner boundaries of the pulsation region the eigenvalues of the matrix \(M\), Eq.~(\ref{eq:matrixM}), form a large-magnitude pair together with two smaller-magnitude eigenvalues: one of intermediate magnitude and one of smallest magnitude. We thus propose a classification in which the large-magnitude eigenvalue with \(\Im(k)>0\) is denoted by 1, and the large-magnitude one with \(\Im(k)<0\) by 2. The intermediate eigenvalue is assigned index 3 and the smallest-magnitude eigenvalue is assigned index 4. 

Since at each boundary we need to select two out of the four, we introduce the selector notation
\begin{equation}
   (a,b)\text{--}(c,d)
\label{eq:defSelectors}
\end{equation}
to denote the indices of the eigenvalue branches selected for the boundary condition at the surface \((a,b)\) and those at the inner boundary \((c,d)\). In what follows we will use this notation to denote selected forms of dispersion-relation-based boundary conditions. 

Thus, for a given selector, the boundary conditions are constructed from the eigenvectors associated with the selected branches. This yields families of solutions that can be compared across different selector choices in a consistent manner.

\subsection{Criteria for admissible solutions}

Near the inner boundary nonphysical solutions with increasing amplitudes may arise. This is a well known effect (see Dziembowski, 1977) caused by numerical instability of the solution due to the divergent nature of the two solutions integrated from the surface downward and poor selection of boundary conditions. It can be remedied by using WKB approximations but we aim to obtain formulation of boundary conditions enabling integration without the need to resort to approximations. We have therefore adopted selector-ranking criteria based on the value of \(|d|\) and \(|ED|\) at the inner boundary. The rationale for these criteria is that near inner boundary the solutions of Eq.~(\ref{eq:basicEq}) may be described (WKB) as: \(d\sim s/k_T\), \(p\sim s/k_T^2\) that is, both the displacement and pressure perturbations become asymptotically small compared to the entropy perturbation. And since \(|d|\gtrsim |p|\) we use only the value of displacement perturbation. On the other hand the behavior of dissipation integral near the inner boundary is a sensitive indicator of the behavior of the eigenmode. When the eigenmode does not follow WKB asymptotics \(|ED|\) tends to diverge. 

While additional diagnostics (e.g. contamination coefficients) are also considered, the above criteria are sufficient for verifying optimal selectors.

\subsection{Reduction of possible selectors}

For three models with \(\log(T_{eff})=3.8,\ 3.85,\ 3.9\), representative of the central part of the temperature range considered here, we selected initial frequencies corresponding to either the fundamental mode or a low-frequency strange mode and performed frequency iteration for all selector combinations. This yielded, for each model, up to 36 candidate solutions corresponding to different selector choices. In some cases the iterations did not converge or converged outside the prescribed tolerance range, such cases were excluded.

Among the remaining solutions, the majority of selector choices led to eigenmodes exhibiting regular behavior near the inner boundary and approaching the expected WKB asymptotics in the deep envelope. This indicates that a broad class of boundary-condition selectors is admissible, provided that the selected subspace suppresses strongly divergent components of the solution.

Within this admissible class, certain selectors have more favorable numerical properties. In particular, the selectors \((2,4)\text{--}(1,3)\) and \((2,3)\text{--}(1,3)\) represent a particularly well-conditioned class of admissible boundary conditions. These selectors share a common structure: at the outer boundary they combine a dominant acoustic branch with a slow (non-acoustic) branch, while at the inner boundary they include a large-magnitude branch together with an intermediate-magnitude one. This structure is consistent with the WKB expectations in the deep envelope.

Other admissible selectors generally produce solutions that also follow the asymptotic behavior, although with somewhat larger values of \(|d|\) near the inner boundary. The differences are therefore quantitative rather than qualitative.

These results suggest that the primary role of the inner boundary condition is to select an admissible subspace what eliminates strongly divergent components of the solution. Once this condition is satisfied, the detailed form of the inner selector has only a minor influence on the resulting eigenfunctions and, as shown in the next section, a negligible effect on the eigenfrequencies.

\subsection{Deep envelope behavior}

\begin{figure}[htb]
	\includegraphics{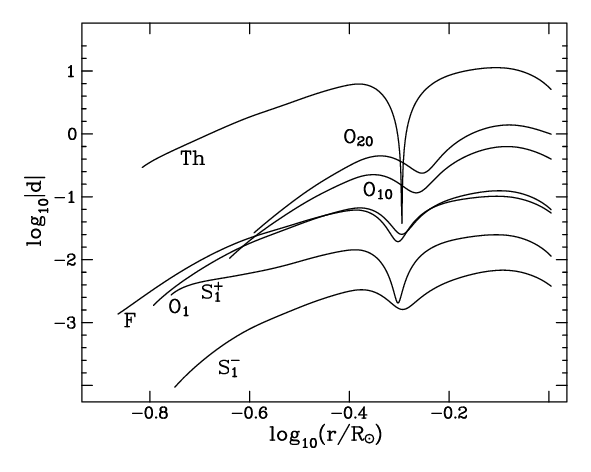}
	\FigCap{Behavior of the amplitude of displacement perturbation deep in the envelope for model \(\log(T_{eff})=3.85\) and boundary conditions selector \((2,4)\text{--}(1,3)\). Shown are modes: \(F\)-fundamental, \(O_{n}\)-overtones, \(S^{+}\)-strange excited, \(S^{-}\)-strange damped and \(Th\)-thermal mode. Below \(\log(r/R_{\odot})\sim -0.6\) the modes enter an exponentially decaying regime consistent with WKB predictions.}
\label{fig:Fig1}	
\end{figure}

In Fig.~\ref{fig:Fig1} the behavior of the amplitude of displacement perturbation deep in the envelope is shown. The regions shown correspond to \(\log(T)\sim 5.5-6.8\). 

From Fig.~\ref{fig:Fig1} it is seen that the modes attain asymptotic behavior in the deep layers of the envelope in agreement with predictions from WKB analysis (see Dziembowski, 1977). 

In the next Section we will be using several of the selectors, not only the one for which results are presented in Fig.~\ref{fig:Fig1} to discuss frequency spectra. For all the selectors considered, the resulting eigenfunctions were verified to remain regular throughout the envelope. While the detailed behavior near the inner boundary may depend on the selected subspace, no spurious divergencies or numerical instabilities were encountered for those selectors, thus the solutions remain well-behaved in the deep layers of the envelope.

These results indicate that, once the solution is aligned with an admissible subspace at the inner boundary the form of the pulsation spectrum is governed primarily by the choice of the outer boundary selectors. This dependence is examined in detail in the following section. 

\section{Dependence of the pulsation spectrum on boundary condition selectors}

In this section we examine the dependence of the pulsation spectrum on the choice of boundary condition selectors. We have performed a systematic analysis and found out from the results that the spectra can be grouped into several distinct categories dependent on the outer boundary selector. We have also found that for a chosen outer boundary selector the effect of the inner boundary selector on the eigenfrequencies is minimal. Thus the spectrum is primarily determined by the outer boundary conditions. Hence it is sufficient to label spectra categories by the outer boundary selector. The results will are presented in the following subsections.

\subsection{Spectra for outer selectors (2,3) and (2,4)}

We start with these selectors as they are close to the often used dispersion-relation-based outer boundary condition supplemented by a blackbody relation. In particular selectors of the form: \((2,m)\text{--}(1,n),\  m,n\in {3,4}\) represent a conventionally adopted dispersion relation at the outer boundary combined with WKB-derived solution at the inner boundary. Thus these two selectors are close to what is being used when setting boundary conditions in case of supergiant envelope pulsations.

It may be noted that a selector of this kind i.e. \((2,m)\text{--}(1,n)\) may be interpreted as corresponding predominantly to outward-propagating components in the dominant channels at both boundaries. A situation expected under normal conditions in case of supergiant envelope pulsations.

\begin{figure}[htb]
	\includegraphics{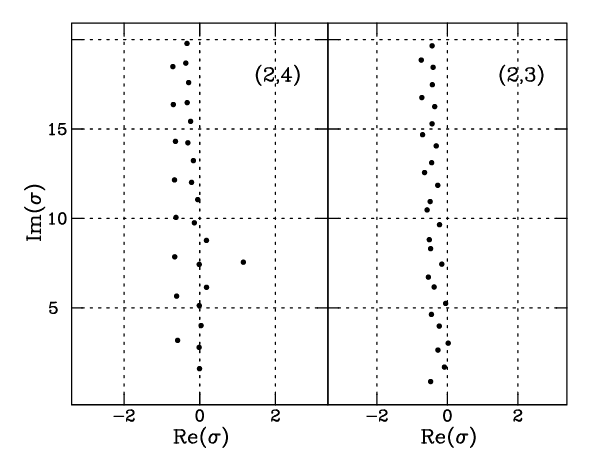}
	\FigCap{Eigenfrequencies for the model \(\log(T_{eff})=3.8\). Left pane - frequencies for selectors \((2,4)\). Right pane - data for selectors \((2,3)\).}
\label{fig:Fig2}	
\end{figure}

In Fig.~\ref{fig:Fig2} the map of eigenfrequencies for boundary conditions selectors of the form \((2,4)\text{--}(a,b)\) and \((2,3)\text{--}(a,b),\  a,b\in \{1,2,3,4\}\) is presented. In what follows we will use the abbreviated notation such as \((2,4)\) to denote selector classes.

We have found that the eigenfrequencies obtained for any inner boundary selector and for a specific outer boundary selector are within, for the \((2,4)\) selector, \(~10^{-3}\) for real part of \(\sigma\), and within \(~10^{-5}\) for imaginary part. That is for any of the possible inner boundary selectors \((a,b),\ a,b\in \{1,2,3,4\}\) the relative shift in eigenfrequency does not exceed the given values. The same holds for the other kinds of outer boundary selectors.

In case of thermal modes, not discussed here, there are changes for particular kinds in inner selectors, but this topic is outside of the scope of current paper.

From Fig.~\ref{fig:Fig2} it may be seen that the spectrum exhibits excited (left pane) and damped (both panes) strange modes. The spectrum depends sensitively on which slow branch (3 or 4) accompanies the dominant acoustic branch at the outer boundary. Interestingly the strongly excited strange mode appears when it is the least-magnitude eigenvalue branch that is adopted. 

This means that when, at the outer boundary, a fast changing (in \(x=\ln(r/R_{\odot})\)) branch is selected with a least changing one a strongly excited strange mode appears. And irrespective of whether the slow branches 3 or 4 are selected to accompany the fast outgoing acoustic like branch there appear damped strange modes, though their damping rates depend on the branches. They are clearly visible in the left pane diagram but are also evident in the one in the right pane.

\subsection{Spectra for outer selectors \((1,3)\) and \((1,4)\)}

While the adoption of an incoming acoustic-like perturbation for the dominant branch at the outer boundary is not a standard choice it is nevertheless interesting to check what kind of spectrum would such a selection lead to. The rationale for considering outer boundary condition of the form \((1,m),\ m\in {3,4}\) is that for dispersion relation based boundary conditions it is easily identifiable and can be avoided, but the same can not be said about other forms of boundary conditions. It should be pointed that with our setup (Eq.~(\ref{eq:obcCond})) the solution vector at the surface spans the subspace determined by the boundary condition.

\begin{figure}[htb]
	\includegraphics{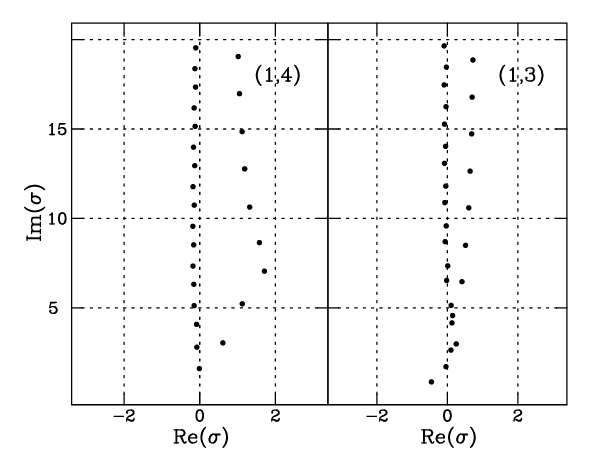}
	\FigCap{Same model as in Fig.~\ref{fig:Fig2}. Left pane - frequencies for selectors \((1,4)\). Right pane - data for selectors \((1,3)\). This choice of outer boundary conditions leads to sequences of excited strange modes with damped strange modes not present.}
	\label{fig:Fig3}	
\end{figure}

In Fig.~\ref{fig:Fig3} the spectra for the two outer boundary selectors are presented. It may be seen the that the choice of incoming acoustic-like perturbation at the outer boundary enhances the excitation of strange modes, leading to sequences dominated by unstable modes, with damped counterparts largely reduced in value. This occurs for both minor branch selections, but for the branch 4 - the smallest-magnitude eigenvalue, the excitation is higher than for the branch 3, similarly to what was seen in Fig.~\ref{fig:Fig2}. The most excited strange mode for branch 4 occurs at similar frequency (\(\Im(\sigma)\sim 7.5\)) as for the selector \((2,4)\). Interestingly for both \((1,3)\) and \((1,4)\) the damped strange modes sequence is non existent. This also is a different behavior to the case where the dominant branch was 2. 

Thus it would seem (at least for the studied model) that the selection of incoming acoustic branch leads to stronger excitation of strange modes than does the selection of the outgoing acoustic branch have, while the minor branch effect is strongest in the case of slowest branch 4. Similarly to the previous cases the spectra for selectors \((1,3)\) and \((1,4)\) are not affected by the choice of any of the inner boundary subspace.

\subsection{Spectrum for outer selectors \((1,2)\)}

The outer boundary selectors of the form \((1,2)\) represent a 2D subspace at the outer boundary made from vectors spanning the two acoustic-like branches. Thus the dynamics at near the boundary is determined by two fast (large eigenvalue \(k\)) branches.

\begin{figure}[htb]
	\includegraphics{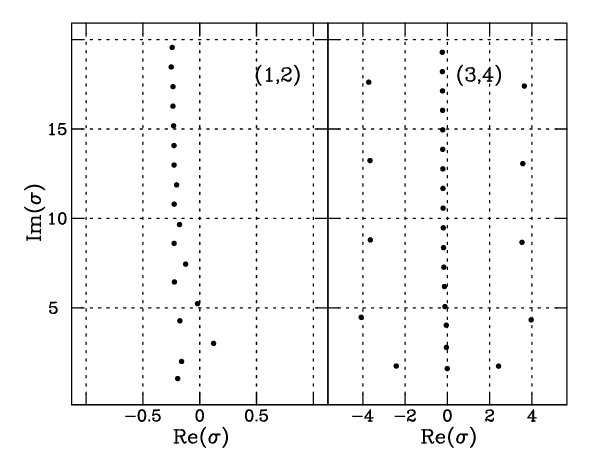}
	\FigCap{Same model as in Fig.~\ref{fig:Fig2}. Left pane - frequencies for selectors \((1,2)\). Right pane - data for selectors \((3,4)\). Please note that the scales on x-axis in both panes differ substantially and they differ from scales on previous plots. The \((1,2)\) form of outer boundary conditions leads to a acoustic modes like spectrum. The boundary conditions selector \((3,4)\) leads to two branches of very strongly excited/damped strange modes and a normal looking p-mode branch.}
	\label{fig:Fig4}	
\end{figure}

In the case of the \((1,2)\) selector the frequencies in the spectrum below \(\Im(\sigma) < 10\) seem to bifurcate, possibly indicating a more complicated dynamics hidden in the low frequency region.

The outer boundary conditions of the form \((1,2)\) may also be viewed as a generalization of the reflective boundary condition. It could approach such form if the relative contributions of the two independent solutions of Eq.~(\ref{eq:basicEq}) were comparable. 

By comparing the effect on the spectrum of an outer boundary selector of the form \((1,2)\), i.e. made up of two fast branches to the spectra for the outer selectors based on a mixture of one fast branch and one slow branch it may be concluded that when the solution of Eq.~(\ref{eq:basicEq}) near boundary is initiated with two fast eigenvectors of \(M\) then the strange modes are not apparent in the spectrum, while for a mixture of one fast-one slow branches they become evident in the spectrum.

\subsection{Spectrum for outer selectors \((3,4)\)}

The subspace obtained from the eigenvectors of \((3,4)\) at the outer boundary corresponding to the slow, small-magnitude eigenvalues of Eq.~(\ref{eq:matrixM}) represents a substantially different dynamics at the boundary than for the previously examined cases.

The spectrum for the slow branches \((3,4)\) shown in the right pane in Fig.~\ref{fig:Fig4} does contain a p-mode ladder which is clearly separated from the two branches of highly excited and damped strange modes. It should be noted that the resulting spectrum exhibits strongly excited and strongly damped strange modes, with growth and damping rates significantly exceeding those obtained for mixed selectors.

This suggests that allowing the solution near outer boundary to occupy a subspace dominated by slow branches enhances the effective coupling between acoustic and entropy components and leads to more efficient excitation or damping for strange modes.

\subsection{Spectrum for standard boundary conditions}

For the same model we have computed spectrum using typically adopted boundary conditions of \(p'=0\) and linearized black-body relation at the outer boundary, and an inner boundary condition based on WKB solutions (Dziembowski, 1977) of Eq.~(\ref{eq:basicEq}). 

\begin{figure}[htb]
	\includegraphics{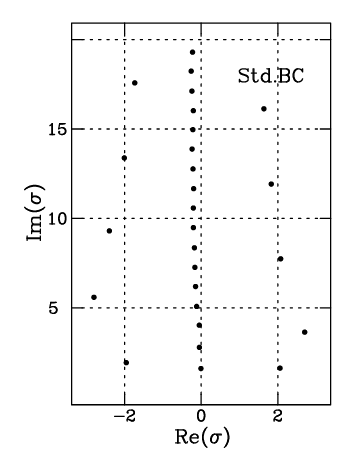}
	\FigCap{Same model as in Fig.~\ref{fig:Fig2}. The spectrum of frequencies for the classical boundary conditions.}
	\label{fig:Fig5}	
\end{figure}

The frequency spectrum for this form of classical boundary conditions is presented in Fig.~\ref{fig:Fig5}. As may be seen from this figure that the obtained spectrum exhibits both normal and strange modes and the strange modes are arranged symmetrically relative to \(\Re(\sigma)=0\) axis.

The spectrum resembles the one presented in the right pane of Fig.~\ref{fig:Fig4} for selectors \((3,4)\). This indicates that this form of outer boundary conditions allows the solution to occupy a subspace close to that spanned by the slow branches. But it should be noted that the excitation rates for the strange modes obtained with our classical boundary conditions are somewhat smaller than for the \((3,4)\) selectors. Which indicates an admixture of a fast branch into the solution at the boundary.

We have projected the solution onto the eigenvectors of \(M\) at the surface using \(\gamma\)'s given by Eq.~(\ref{eq:defGamma}) for representative strange modes. We have found that these modes have largest projection onto the subspace spanned by slow branches. For example, the normalized projection of the strongest excited strange mode from Fig.~\ref{fig:Fig5} was \((0.09, 0.09, 1, 0.9)\). The ordinary p-modes showed similar dominance of the same components. This supports the interpretation that, at least for the studied model, the standard boundary conditions admit a surface subspace close to that spanned by branches \((3,4)\), consistent with the similarity between Fig.~\ref{fig:Fig5} and the \((3,4)\) spectrum in Fig.~\ref{fig:Fig4}.

The spectrum shown in Fig.~\ref{fig:Fig5} may be also compared with the three spectra presented in Zalewski (1992) on Fig.~7 where frequencies of pulsation for the classical outer boundary conditions, the dispersion relation (as formulated in the cited paper) and the envelope with radiative transfer are presented for a model with the same parameters as adopted here. In Fig.~7 it is seen that the strange modes exist for classical boundary conditions while their excitation rates are diminished for the other forms of boundary conditions. For example the lowest frequency excited strange mode seen in Fig.~7 disappears while higher order excited strange modes excitation is decreased. 

Within the present classification of spectra by outer boundary selectors, the behavior reported in Zalewski (1992) may be interpreted as a consequence of the degree to which the adopted boundary conditions permit coupling between acoustic and entropy components. More generally, this suggests that variations in strange modes excitation reported in earlier studies may be understood in terms of how strongly the boundary conditions permit interaction between these components.
 
\section{Conclusions}

We have formulated boundary conditions for radial pulsations in AGB/\allowbreak post-AGB envelopes in terms of subspaces spanned by eigenvectors of the local dispersion relation. This approach provides a systematic framework for constructing and comparing admissible boundary-condition selectors.

It is found that the inner boundary condition primarily acts as a selector of admissible solutions, eliminating divergent components and ensuring regular behavior in the deep envelope. In contrast, the outer boundary condition determines the character of the pulsation spectrum. For a given outer selector, variations of the inner selector have only a minor effect on the resulting eigenfrequencies and growth rates.

The pulsation spectra can be classified according to the pair of branches selected at the outer boundary. Selectors involving both acoustic and slow branches give rise to classical strange-mode spectra with moderate growth and damping rates. Selectors composed of two acoustic branches produce spectra resembling p-modes, with strange modes not prominent within the explored frequency range. In contrast, selectors based on two slow branches lead to strongly excited and strongly damped strange modes, forming nearly symmetric pairs in the complex frequency plane.

These results suggest that the role of the outer boundary condition is to determine the relative contribution of acoustic and slow components in the solution. In this sense, the boundary condition acts as a selector of admissible subspaces rather than as a direct source of physical forcing. The degree of strange-mode excitation is then controlled by the extent to which the selected subspace allows interaction between acoustic and entropy components.

This interpretation provides a unified view of the dependence of strange-mode spectra on boundary conditions and offers a natural explanation of earlier results obtained using classical boundary-condition formulations.

We have also verified that the same classification of spectra by outer boundary selector persists for different effective temperatures for the studied model, although the detailed morphology of the spectra varies.

\end{document}